\documentclass[twocolumn,showpacs,preprintnumbers,amsmath,amssymb,superscriptaddress]{revtex4}

\usepackage{graphicx}
\usepackage{epstopdf}
\usepackage{dcolumn}
\usepackage{bm}

\begin{document}
　
\title{Knight shift vs hole concentration in Hg1201 and Hg1212}

\author{Y. Itoh}
 \affiliation{Department of Physics, Graduate School of Science, Kyoto Sangyo 
University, Kamigamo-Motoyama, Kita-ku Kyoto 603-8555, Japan} 　

\date{\today}

\begin{abstract}
We studied the hole concentration dependences of $^{63}$Cu Knight shifts in single-CuO$_2$-layer high-$T_\mathrm{c}$ cuprate superconductors HgBa$_2$CuO$_{4+\delta}$ and double-layer HgBa$_2$CaCu$_2$O$_{6+\delta}$. We found that the spin Knight shift at room temperature as a function of the hole concentration in the single-layer superconductor is different from that in the double-layer superconductor. Two type relations between the spin Knight shift and the hole doping level serve to estimate the individual hole concentrations of the non-equivalent CuO$_2$ planes in a unit cell.    
\end{abstract}


\maketitle
 
\section{Introduction} 
The doped hole carriers have been believed to distribute nonuniformly in the multi-layer cuprate superconductors. How to estimate the individual carrier doping levels of the non-equivalent CuO$_2$ planes in the multi-layer superconductors has been an issue. The bond valence sum associated with a local ionic valence has served to estimate the hole concentration~\cite{Cava,Brown,Tokura}. The bond valence sum is a microscopic characterization based on the nanometer-scale structural analysis of the ionic bonds. Although a bulk observable such as thermoelectric power (Seebeck coefficient) at room temperature has been proposed to estimate the hole concentration~\cite{Tallon}, no macroscopic observables enable us to estimate the layer-selective hole concentration in the multi-layer systems. Then, microscopic observables have been desired. 

The $ab$-plane spin component of the plane-site $^{63}$Cu Knight shift $^{63}K_s^{ab}$ at room temperature was used to estimate the individual hole concentrations of the non-equivalent CuO$_2$ planes in the multi-layer superconductors~\cite{Mukuda1, Mukuda2}. The hole concentration $p$ is defined by the fraction of holes per Cu$^{2+p}$ ion in the CuO$_2$ plane. A one-to-one correspondence between $^{63}K_s^{ab}$ and the hole concentration $p$ has been proposed except for La$_{2-x}$Sr$_x$CuO$_4$ (LSCO), YBa$_2$Cu$_3$O$_{7-\delta}$ (Y1237) and YBa$_2$Cu$_4$O$_8$ (Y1248)~\cite{Mukuda1,Mukuda2}. The single universal relation between $^{63}K_s^{ab}$  and $p$ has bee adopted for the multi-layer superconductors. However, the uniform spin susceptibilities per Cu spin in single-layer systems are known to be larger than those in double-layer systems~\cite{Millis1, Millis2}. It should be noted that the difference in $^{63}K_s^{ab}$ of single-CuO$_2$-layer superconductor HgBa$_2$CuO$_{4+\delta}$ (Hg1201) and double-layer HgBa$_2$CaCu$_2$O$_{6+\delta}$ (Hg1212) in~\cite{Itoh1, Itoh2, Itoh3} has been overlooked. Thus, an issue how relevant the application of the single universal relation between $^{63}K_s^{ab}$  and $p$ is for the multi-layer systems should be addressed.  
 
\begin{figure}[h]
 \begin{center}
 \includegraphics[width=0.80\linewidth]{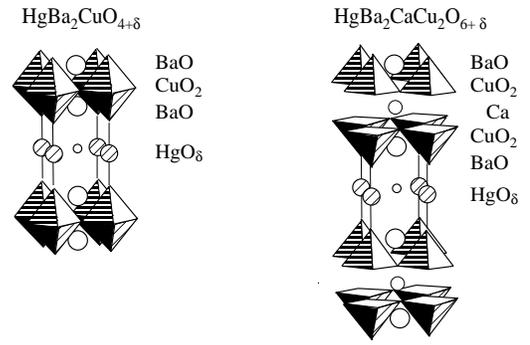}
 \end{center}
 \caption{\label{fig:CuNQR}
Schematic crystal structures of Hg1201 (left) and Hg1212 (right). 
 }
 \end{figure}
 
The optimal $T_\mathrm{c}$’s of Hg1201 and Hg1212 are the highest among the single-layer and the double-layer systems. The structural flatness of the CuO$_2$ plane characterizes Hg1201 and Hg1212. Figure 1 shows the schematic crystal structures of Hg1201 and Hg1212. The oxygen concentration in the HgO layer controls the hole doping level in the CuO$_2$ plane. The crystal symmetry is tetragonal. We believe that Hg1201 and Hg1212 are the canonical high-$T_\mathrm{c}$ cuprate superconductors.

In this paper, we studied the hole concentration dependences of the spin Knight shifts at room temperature in the single-layer superconductors Hg1201 and the double-layer Hg1212 in~\cite{Itoh1, Itoh2, Itoh3}. The present analysis is based on the NMR results reported in~\cite{Itoh1, Itoh2, Itoh3}. The preparation methods of the samples for the NMR measurements are described in~\cite{Itoh1, Itoh2, Fukuoka}. We emphasize that the spin Knight shift as a function of the hole concentration in the single-layer superconductors is different from that in the double-layer superconductors

\section{Experimental Knight shifts of Hg1201 and Hg1212} 
\begin{figure*}[t]
 \begin{center} 
\includegraphics[width=0.70\linewidth]{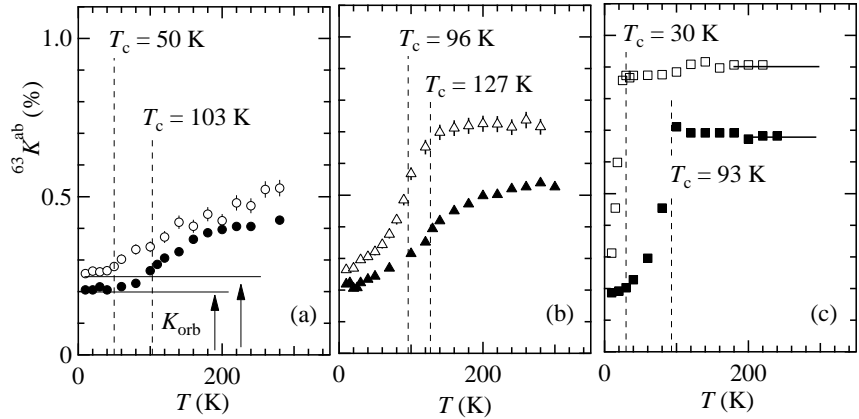}
 \end{center}
 \caption{\label{fig:rec}
$^{63}$Cu Knight shifts $^{63}K^{ab}$ for Hg1201 (open symbols) and Hg1212 (closed symbols) in the underdoped (a), the optimally doped (b), and the overdoped regimes (c), which are reproduced from~\cite{Itoh1, Itoh2, Itoh3}. The dashed lines indicate the individual $T_\mathrm{c}$’s. The sold lines in (c) are visual guides.    
 }
\end{figure*} 

Figure 2 shows the $^{63}$Cu Knight shifts $^{63}K^{ab}$'s of Hg1201 (open symbols) and Hg1212 (closed symbols) in an external magnetic field along the $ab$ planes, which are reproduced from~\cite{Itoh1, Itoh2, Itoh3}. The doping levels are categorized into underdoping (a), optimally doping (b), and overdoped doping (c). 

The $^{63}$Cu Knight shift $^{63}K^{ab}$ is given by $^{63}K^{ab}$ = $^{63}K_s^{ab}$($T$) +$^{63}K_{orb}^{ab}$, where $^{63}K_s^{ab}$ is the spin Knight shift and $^{63}K_{orb}^{ab}$ is the orbital shift. $^{63}K_s^{ab}$ is proportional to the uniform spin susceptibility multiplied by the hyperfine coupling constant. $^{63}K_{orb}^{ab}$ is proportional to the Van Vleck orbital susceptibility. The temperature dependence of $^{63}K^{ab}$ in the cuprate superconductors comes from that of the spin shift $^{63}K_s^{ab}$. We have estimated $^{63}K_{orb}^{ab}$ $\sim$ 0.25 $\%$  for Hg1201 and $\sim$ 0.20 $\%$ for Hg1212~\cite{Itoh1, Itoh2, Itoh3}. As seen in Fig. 2, $^{63}K_s^{ab}$ of Hg1201 is larger than that of Hg1212 at each doping regime.  

Figure 3 shows the hole concentration $p_h$ in~\cite{Fukuoka} against $^{63}K_s^{ab}$ ($\%$) at room temperature of Hg1201 and Hg1212 in~\cite{Itoh1, Itoh2, Itoh3}. From the least squares fits, we obtained two relations of S1: $p_h$ = 0.63$^{63}K_s^{ab}$(RT) - 0.13 (Hg1201) and D1, $p_h$ = 0.74$^{63}K_s^{ab}$(RT) - 0.04 (Hg1212). The solid lines in Fig. 3 indicate the fit functions of S1 and D1. The extrapolations of S1 and D1 to $p_h$ = 0 lead to $^{63}K_s^{ab}$ = 0.21 $\%$ (Hg1201) and $^{63}K_s^{ab}$ = 0.05 $\%$ (Hg1212) at the phase boundary. The empirical functions of S1 and D1 are different from the previous fit functions F1 and F2 ($^{63}K_s^{ab}$ $<$ 0.5 $\%$) adopted for the multi-layer systems in~\cite{Mukuda1}. 
 
\begin{figure}[h]
\begin{center}
\includegraphics[width=0.8\linewidth]{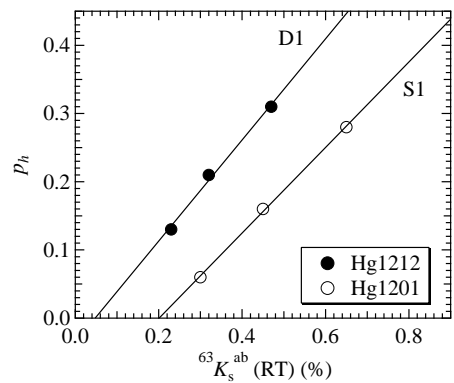}
\end{center}
\caption{\label{fig:Tdep}
Hole concentration $p_h$ against spin Knight shift $^{63}K_s^{ab}$ at room temperature (RT) in Hg1201 and Hg1212. The solid lines are the fit functions of S1: $p_h$ = 0.63$^{63}K_s^{ab}$(RT) - 0.13 (Hg1201) and D1, $p_h$ = 0.74$^{63}K_s^{ab}$(RT) - 0.04 (Hg1212).   
}
\end{figure}  
 
\section{Discussions} 
We discuss the hole concentration dependence upon the spin Knight shift for the other cuprate superconductors and the alternative estimation of the hole concentration. 

\begin{figure*}[t] 
\begin{center}
\includegraphics[width=0.70\linewidth]{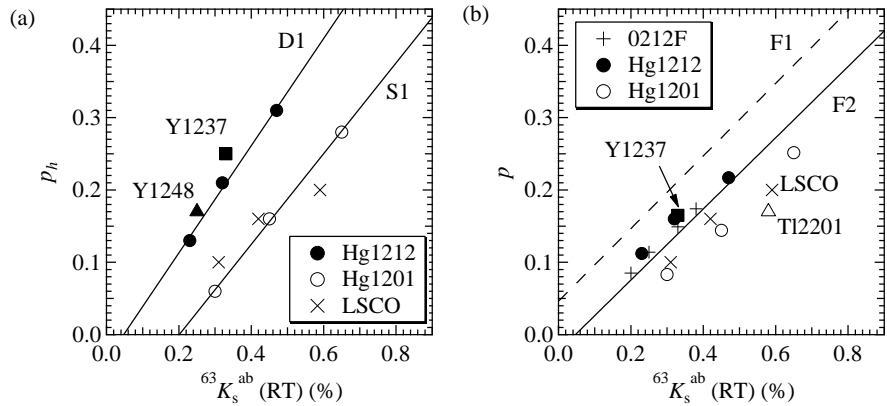}
\end{center}
\caption{\label{fig:tau1}
(a) Hole concentration $p_h$ against spin Knight shift $^{63}K_s^{ab}$ of Hg1201, LSCO~\cite{Ohsugi}; Hg1212, Y1237~\cite{Takigawa, Shimizu} and Y1248~\cite{Machi, Zimmermann1, Zimmermann2}. (b) Hole concentration $p$ against spin Knight shift $^{63}K_s^{ab}$ of Hg1201, LSCO~\cite{Ohsugi}, Tl2201~\cite{Kambe}; Hg1212, 0212F~\cite{Mukuda2}. One should note $p_h$ (a) and $p$ (b) estimated from the different ways. In (b), F1 (dashed line) and F2 (sold line) are the previous functions for $^{63}K_s^{ab}$ $<$ 0.5 $\%$ in~\cite{Mukuda1}. 
}
\end{figure*}  
  
Figure 4(a) shows $p_h$ against $^{63}K_s^{ab}$ ($\%$) of Hg1201, LSCO~\cite{Ohsugi}; Hg1212, Y1237~\cite{Takigawa, Shimizu} and Y1248~\cite{Machi, Zimmermann1, Zimmermann2}. We estimated $^{63}K_s^{ab}$ at room temperature for LSCO by linear extrapolation from the existing data in~\cite{Ohsugi}. The single-layer system LSCO is located close to S1. The double-layer systems of Y1237 and Y1248 are located close to the line D1. 

In Fig. 4(b), we estimated the hole concentrations $p$’s by the parabolic curve of $T_\mathrm{c}$/$T_\mathrm{c, max}$ = 1 - 82.6($p$ - 0.16)$^2$ in~\cite{Tallon2} after~\cite{Mukuda1}. Figure 4(b) shows $p$ against $^{63}K_s^{ab}$ ($\%$) of Hg1201, LSCO~\cite{Ohsugi}, Tl$_2$Ba$_2$CuO$_{6+\delta}$ (Tl2201)~\cite{Kambe}; Hg1212, and Ba$_2$CaCu$_2$O$_4$(F,O)$_2$ (0212F)~\cite{Mukuda2}. In spite of the alternative estimation of the hole concentration, the $p$ vs $^{63}K_s^{ab}$ dependence of Hg1201 is different from that of Hg1212. The reason why $^{63}K_s^{ab}$'s of Hg1212 are smaller than those of Hg1201 at the respective doping levels in Figs. 2 and 3 may be due to the effect of the magnetic bilayer coupling. The effect of the bilayer coupling has also been studied for the triple-layer superconductors. Thus, one should take into consideration which type relation is relevant S1 or D1 to estimate the individual hole concentrations of the non-equivalent CuO$_2$ planes in the multi-layer superconductors.  

\section{Conclusion} 

We found that the spin Knight shift $^{63}K_s^{ab}$ as a function of the hole concentration $p_h$ in the single-layer superconductors Hg1201 is different from that in the double-layer superconductors Hg1212. Since we believe that Hg1201 and Hg1212 are the canonical systems, the $p_h$ dependences upon $^{63}K_s^{ab}$ at room temperature in Hg1201 and Hg1212 should be standard to estimate the individual hole concentrations of the non-equivalent CuO$_2$ planes in a unit cell.


\section{References}
 \begin{thebibliography}{50}

\bibitem{Cava}R. J. Cava, A. W. Hewat, E. A. Hewat, B. Batlogg, M. Marezio, K. M. Rabe, J. J. Krajewski, W. F. Peck Jr., L. W. Rupp Jr., Physica C 165 (1990) 419-433.   

\bibitem{Brown}I. D. Brown, J. Solid State Chem. 90 (1991) 155-167. 

\bibitem{Tokura} Y. Tokura, J. B. Torrance, T. C. Huang, A. I. Nazzal, Phys. Rev. B 38 (1988) 7156-7159.

\bibitem{Tallon} J. L. Tallon, C. Bernhard, H. Shaked, R. L. Hitterman, J. D. Jorgensen, Phys. Rev. B 51 (1995)  12911-12914.

\bibitem{Mukuda1} H. Mukuda, S. Shimizu, A. Iyo, Y. Kitaoka, J. Phys. Soc. Jpn. 81 (2012) 011008. 

\bibitem{Mukuda2} S. Shimizu,  S. Iwai, S. Tabata, H. Mukuda, Y. Kitaoka, P. M. Shirage, H. Kito, A. Iyo, Phys. Rev. B 83 (2011) 144523.

\bibitem{Millis1} A. J. Millis, L. B. Ioffe, H. Monien, J. Phys. Chem. Solids, 56 (1995) 1641-1643. 

\bibitem{Millis2} A. J. Millis, H. Monien, Phys. Rev. Lett. 70 (1993) 2810-2813.  

\bibitem{Itoh1} Y. Itoh, T. Machi, S. Adachi, A. Fukuoka, K. Tanabe, H. Yasuoka, J. Phys. Soc. Jpn. 67 (1998) 312-317. 
  
 \bibitem{Itoh2} Y. Itoh, A. Tokiwa-Yamamoto, T. Machi, K. Tanabe, J. Phys. Soc. Jpn. 67 (1998) 2212-2214. 
 
\bibitem{Itoh3} Y. Itoh, T. Machi, in "Superconducting Cuprates: Properties, Preparation and Applications," ed. Koenraad N. Courtlandt (Nova Science Publisher, NY, 2009) p.p. 235 - 268. 
  
\bibitem{Fukuoka} A. Fukuoka, A. Tokiwa-Yamamoto, M. Itoh, R. Usami, S. Adachi, K. Tanabe, Phys. Rev. B 55 (1997) 6612-6620.   

\bibitem{Ohsugi} S. Ohsugi, Y. Kitaoka, K. Ishida, G.-q. Zheng, K. Asayama, J. Phys. Soc. Jpn. 63 (1994) 700-715.  

\bibitem{Takigawa} M. Takigawa, P. C. Hammel, R. H. Heffner, Z. Fisk, J. L. Smith, R. B. Schwarz, Phys. Rev. B 39 (1989) 300-303. 

\bibitem{Shimizu} T. Shimizu, H. Aoki, H. Yasuoka, T. Tsuda, Y. Ueda, K. Yoshimura, K. Kosuge, J. Phys. Soc. Jpn. 62 (1993) 3710-3720. 
 
\bibitem{Machi}T. Machi, I. Tomeno, T. Miyatake, N. Koshizuka, S. Tanaka, T. Imai, H. Yasuoka, Physica C 173 (1991) 32-36.

\bibitem{Zimmermann1} H. Zimmermann, M. Mali, I. Mangelschots, J. Roos, L. Psuli, D. Brinkmann, J. Karpinski, S. Rusiecki, E. Kaldis, J. Less-Commom Metals, 164-165 (1990) 138-145. 

\bibitem{Zimmermann2} H. Zimmermann, M. Mali, M. Bankay, D. Brinkmann, Physica C 185-189 (1991) 1145-1146. 

\bibitem{Tallon2} M. R. Presland, J. L. Tallon, R. G. Buckley, R. S. Liu, N. E. Flower, Physica C 176 (1991) 95-105.

\bibitem{Kambe} S. Kambe, H. Yasuoka, A. Hayashi, Y. Ueda, Phys. Rev. B 47 (1993) 2825-2834. 

\end{thebibliography}
\end{document}